\documentclass[prl,twocolumn,english,showpacs]{revtex4-1}
\usepackage{graphicx}
\usepackage{amssymb}

\begin{document}

\title{Radiofrequency spectroscopy of a strongly interacting two-dimensional Fermi gas}

\author{Bernd Fr{\"o}hlich$^1$, Michael Feld$^1$, Enrico Vogt$^1$, Marco Koschorreck$^1$, Wilhelm Zwerger$^2$, and Michael K{\"o}hl$^1$}

\affiliation{$^1${Cavendish\,Laboratory,\,University~of Cambridge, JJ Thomson Avenue, Cambridge CB30HE, United Kingdom}\\$^2${Technische\,Universit{\"a}t\,M{\"u}nchen,\,Physik\,Department,\,James-Franck-Strasse,\,85748 Garching,\,Germany}}

\begin{abstract}
We realize and study a strongly interacting two-component atomic Fermi gas confined to two dimensions in an optical lattice. Using radio-frequency spectroscopy we measure the interaction energy of the strongly interacting gas. We observe the confinement-induced Feshbach resonance on the attractive side of the 3D Feshbach resonance and find the existence of confinement-induced molecules in very good agreement with theoretical predictions.
\end{abstract}

\pacs{03.75.Ss 
05.30.Fk 
68.65.-k 
}

\date{\today}

\maketitle

Two-dimensional Fermi gases play a pivotal role in quantum many-body physics. The restriction of particle motion to a plane profoundly increases the role of fluctuations and leads to qualitatively new effects in the interparticle interaction \cite{Randeria1989,Petrov2000a,Petrov2001,Petrov2003,Wouters2003,Naidon2007,Bloch2008}. In the solid state context, strongly interacting two-dimensional Fermi gases are found in the cuprates, the two-dimensional electron gas in nanostructures, and in thin $^3$He films. With the advent of ultracold atomic Fermi gases \cite{DeMarco1999c} and the ability to confine them to two-dimensional configurations \cite{Modugno2003,Jochim2003b,Gunter2005,Martiyanov2010,Dyke2010}, research has revived because tunable and ultraclean samples have become available. The direct experimental access to microscopic parameters, such as the particle interaction, promotes ultracold two-dimensional Fermi gases as quantum simulators of fundamental many-body effects.

A quantum system is kinematically two-dimensional if the chemical potential and the thermal energy are smaller than the energy gap $\hbar \omega$ to the first excited state in the strongly confined direction. For harmonic confinement, the motion of particles is restricted to the quantum mechanical ground state with an extension $l_0=\sqrt{\hbar/m \omega}$, in which $m$ is the mass of the particles. This new length scale $l_0$ competes with the three-dimensional s-wave scattering length $a$. As a result, two-dimensional gases display features not encountered in their three-dimensional counterparts. Beyond the absence of a true condensate at finite temperature there is also no scale invariant regime similar to the unitary gas at infinite scattering length. This has to do with the peculiar features of two-body scattering.
Specifically, the amplitude of the outgoing cylindrical wave for low energy scattering with relative momentum $q$ in two dimensions is of the form \cite{Petrov2001,Petrov2003,Bloch2008}
\begin{equation}
f(q)=\frac{4\pi}{\ln(1/q^2 a_{2D}^2)+i\pi}\, \label{coupling},
\end{equation}
which defines the 2D scattering length $a_{2D}$. The logarithmic dependence on momentum shows that $f(q)$ is never independent of energy and indicates that the dimensionless interaction strength $1/\ln(k_Fa_{2D})$ depends logarithmically on the Fermi wave vector \cite{Bloom1975,Bertaina2010}. In particular, the weak coupling limit of a Fermi gas, whose interactions can be described in a mean-field picture, is reached at {\it high} densities.

Two-dimensional confinement also stabilizes a bound dimer state, the presence of which is both a necessary and also a sufficient criterion for the existence of an s-wave pairing instability in 2D  \cite{Randeria1989}. For weak attractive interactions, with a negative scattering length and $|a|<l_0$, the dimer binding energy is predicted to be \cite{Bloch2008}
\begin{equation}
E_B=0.905\left(\hbar\omega/\pi\right)\,\exp\left(-\sqrt{2\pi}\l_{0}/|a|\right)\, .
\label{ebinding}
\end{equation}
The associated size of the dimer is related to the two-dimensional scattering length $a_{2D}=\hbar/\sqrt{mE_B}$, which is always positive. A weakly bound dimer state with $E_B\ll\hbar\omega$ therefore corresponds to a system in the mean field regime $k_Fa_{2D}\gg 1$, quite in contrast to the situation in three dimensions, where $k_Fa\gg 1$ is the unitary regime of strongest interactions. Near the Feshbach resonance, where $a\to -\infty$, Equation (2) for the bound state energy no longer applies and is replaced by $E_B(a=\infty)=0.244\hbar\omega$ \cite{Bloch2008}. The binding energy is thus a universal constant in units of the transverse confinement energy, similar to the case where the atoms are confined in one dimensional tubes \cite{Bergeman2003,Moritz2005}. This prediction will be verified experimentally below. On the side $a>0$ of the Feshbach resonance, where a two-body bound state already exists in three dimensions, the binding energy eventually approaches the standard 3D result $E_B=\hbar^2/ma^2$ because binding is unaffected by the existence of a confinement potential once the size of the dimers becomes smaller than the characteristic length $l_0$. In the many-body system, this is the BEC limit, where $k_Fa_{2D}\ll 1$. The crossover between both limits occurs at intermediate coupling $k_Fa_{2D}\simeq 1$, where the two-body scattering amplitude (1) is purely imaginary.

In this Letter, we report on the realization of a strongly interacting two-dimensional spin-1/2 Fermi gas and radio-frequency (rf) spectroscopy to determine its energy spectrum. We measure the interaction energy of the two-dimensional gas and observe the confinement-induced scattering resonance. Furthermore, we detect the confinement-induced bound states in an attractively interacting Fermi gas. Previously, only in one-dimensional quantum gases such bound states and resonances have been observed by spectroscopy of Fermi gases \cite{Moritz2005} and loss measurements in bosonic gases \cite{Haller2010}, respectively. In two dimensions, experiments have remained inconclusive because atom loss data have hinted at a confinement-induced resonance on the repulsive side of a three-dimensional Feshbach resonance \cite{Haller2010}, contrary to theoretical predictions \cite{Randeria1989,Petrov2000a,Petrov2001,Petrov2003,Wouters2003,Naidon2007}.

We prepare a quantum degenerate Fermi gas of $^{40}$K atoms in a single species apparatus. Starting from a vapor cell magneto-optical trap containing $5 \times10^8$ atoms, we trap a mixture of the $|m_F=9/2\rangle$ and $|m_F=7/2\rangle$ magnetic sublevels of the $|F=9/2\rangle$ hyperfine ground state in a magnetic quadrupole trap. Atoms in this trap are transported mechanically into an ultrahigh-vacuum chamber, where they are transferred into a magnetic Ioffe-Pritchard trap. After radiofrequency-induced evaporative cooling to a temperature of $\sim 10\,\mu$K we load the atoms into an optical dipole trap formed by two crossed laser beams of 1064\,nm wavelength. The horizontal laser beam propagates along the x-direction and is focussed to an elliptical waist of $w_z=17\,\mu$m and $w_y=65\,\mu$m along the vertical and the horizontal directions, respectively. The second laser beam propagates at an angle of $45^\circ$ with respect to gravity in the yz-plane and has a waist of $170\,\mu$m in the horizontal x-direction and $72\,\mu$m along the orthogonal axis. In the optical dipole trap we transfer the atoms into a 50/50 mixture of the $|m_F=-9/2\rangle\equiv |-9/2\rangle$ and $|m_F=-7/2\rangle\equiv |-7/2\rangle$ states using radiofrequency sweeps and pulses. By continuously lowering the depth of the optical trap we perform evaporative cooling of the atoms until we reach $T/T_{F,3D}\approx 0.2$ with approximately 50000 atoms per spin state where $T_{F,3D}$ denotes the Fermi temperature in three dimensions.

In order to study the Fermi gas in a two-dimensional potential well we employ an optical lattice. The optical lattice beam is formed by a retro-reflected laser beam of wavelength $\lambda=1064$\,nm, focussed to a waist of 140\,$\mu$m and propagating horizontally along the y-axis. We increase the laser power over a period of 200\,ms to reach a final potential depth of up to $V_{lat}=83\,E_{rec}$ which is calibrated by intensity modulation spectroscopy. $E_{rec}=h^2/(2 m \lambda^2)$ is the recoil energy. The trapping frequency along the strongly confined direction is $\omega=2\sqrt{V_{lat} E_{rec}}/\hbar$ which for $V_{lat}=83\,E_{rec}$ is $\omega= 2\pi \times 80\,$kHz. Using adiabatic mapping of the quasi-momentum states \cite{Greiner2001b,Kohl2005b}, we verify that the atoms are loaded into the lowest band of the optical lattice. After loading the optical lattice, we adiabatically reduce the power of the optical dipole trap such that the atoms are confined only by the Gaussian intensity envelope of the lattice laser beams. The radial trapping frequency of the two-dimensional gases is $\omega_\perp=2\pi\times 139$\,Hz for $V_{lat}=83\,E_{rec}$ and we confine approximately $2\times10^3$ atoms per two-dimensional gas at the center of the trap. Along the axial direction we populate approximately 30 layers of the optical lattice potential.

\begin{figure}[htbp]
\includegraphics[width=.75\columnwidth,clip=true]{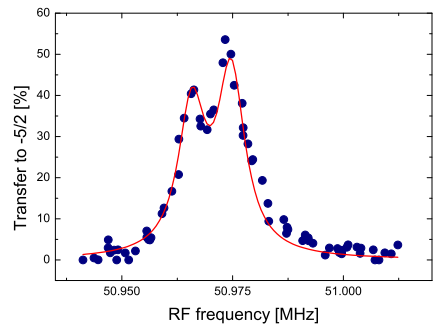}
  \caption{(Color online) Radiofrequency spectrum showing the characteristic two-peak structure of the atomic peak (right), shifted by the interaction energy, and the molecular peak (left). The solid line shows a double Lorentzian fit to extract the peak positions. The spectrum is recorded at 225\,G in an optical lattice of $83\,E_{rec}$ depth.}
 \label{fig1}
\end{figure}

We measure the energy spectrum of a strongly interacting spin-1/2 Fermi gas using radiofrequency spectroscopy. Variations of this technique have proven highly successful in the investigation of three-dimensional \cite{Regal2003a,Regal2003b,Gupta2003,Chin2004,Shin2007,Stewart2008,Baym2007,Hausmann2009} and one-dimensional \cite{Moritz2005} Fermi gases. We ramp the magnetic field close to the Feshbach resonance of the $|-9/2\rangle$ and $|m_F=-5/2\rangle\equiv |-5/2\rangle$ mixture at 224.2\,Gauss. Using a short (100 $\mu$s), rectangular-shaped rf pulse we transfer the atoms from the $|-7/2\rangle$ to the $|-5/2\rangle$ state, which is strongly interacting with the $|-9/2\rangle$ atoms. There is no interaction energy shift between the $|-7/2\rangle$ and $|-5/2\rangle$ states \cite{Gupta2003,Gibble2009}. The magnetic field is calibrated by spin-rotation on a $|-9/2\rangle/|-7/2\rangle$ mixture which also does not experience an interaction shift. Directly after applying the rf pulse, we switch off the optical lattice and apply an inhomogeneous magnetic field to separate the spin components in a Stern-Gerlach experiment. The spatially separated spins are detected by absorption imaging after 7.4\,ms of ballistic expansion. Figure \ref{fig1} shows an rf spectrum of a two-dimensional Fermi gas. It displays a double-peak feature, representing a peak near the atomic Zeeman transition and a peak corresponding to the rf-induced association of molecules.

\begin{figure}[htbp]
  \includegraphics[width=0.8\columnwidth,clip=true]{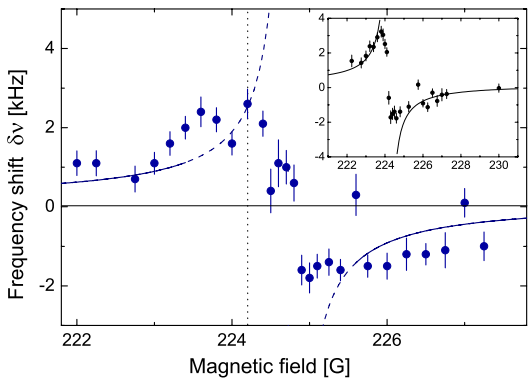}
  \caption{(Color online) Measurement of the interaction energy shift of a two-dimensional Fermi gas in an optical lattice of 83\,$E_{rec}$. The data points show the measured frequency shift as compared to the atomic Zeeman energy. The solid line shows the expected energy shift in the weak coupling regime, the dashed line shows its continuation into the strongly interacting regime. The vertical line indicates the position of the three-dimensional Feshbach resonance at $224.20\pm0.03$\,G. We determine the confinement-induced resonance at $224.72\pm0.05$\,G. The inset shows the measurement for a three-dimensional Fermi gas. Error bars indicate the fitting errors in determining the peak of the spectrum.}
  \label{fig2}
\end{figure}

The interactions of the spin components near the confinement-induced resonance shift the frequency of the atomic rf spin-flip transition from the Zeeman energy. In the weak coupling regime, the interaction energy shift is determined by the coupling constant $g(q)=-\frac{\hbar^2}{m}\frac{2 \pi}{\ln(q a_{2D})}$ \cite{Bloom1975}. In Fig. \ref{fig2} we show the measured interaction energy of a two-dimensional Fermi gas confined to an optical lattice of $83\,E_{rec}$ depth. One identifies a shift of the zero-crossing compared to a three-dimensional Fermi gas (see inset). The solid line shows the theoretical prediction in the weak-coupling regime \cite{Bloom1975}, which for a single two-dimensional layer is $E_{int}/N_5=E_F/[6\ln(\tilde{q} a_{2D})]$. $N_5$ is the number of atoms transferred into the $|-5/2\rangle$ state. $E_F=\sqrt{2N}\hbar \omega_\perp$ is the Fermi energy of $N$ particles per spin state in two dimensions with transverse confinement frequency $\omega_\perp$ which has the value $E_F=h\times 9$\,kHz for our center layer. $\tilde{q}$ is calculated from the density-weighted average of $\hbar^2 q^2/m=2[E_F-m \omega_\perp (x^2+z^2)/2]$ over the Thomas-Fermi profile of a single two-dimensional gas. We average the interaction shift over the 30 layers assuming a Thomas-Fermi envelope of the peak densities along the lattice direction.

\begin{figure}[htbp]
  \includegraphics[width=0.9\columnwidth,clip=true]{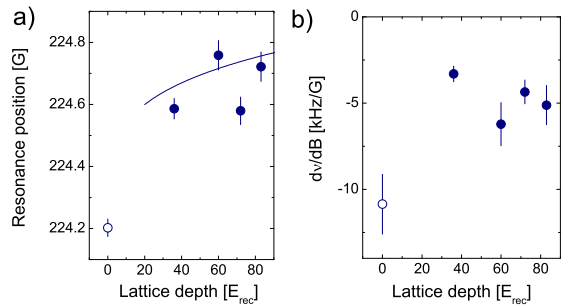}
  \caption{(Color online) (a) Location of the confinement-induced Feshbach resonance vs depth of the optical lattice. The solid line is the prediction of equation (\ref{coupling}) without free parameters. (b) Slope of the interaction energy at the zero-crossing of the confinement-induced resonance. The data points at zero lattice depth in both graphs show the result for the three-dimensional Feshbach resonance. The value of $224.20\pm0.03$\,G is in excellent agreement with \cite{Regal2003b}.}
  \label{fig3}
\end{figure}

We determine the position of the confinement induced resonance by fitting a linear function to the data, from the maximum to the minimum measured value, and determine the zero-crossing of this fit (see Figure \ref{fig3}a). We find a shift of the confinement induced resonance from the location of the three-dimensional Feshbach resonance, which is indicated as the data point at zero lattice depth. The confinement-induced resonance is positioned at negative values of the three-dimensional scattering length, as predicted in \cite{Petrov2001}, but in contrast to loss measurements which located it on the repulsive side of the three-dimensional Feshbach resonance \cite{Haller2010}. The contribution of the initial weakly interacting Fermi gas to the shift of the rf spectrum is negligible since $1/\ln(k_F a_{2D}) \approx 0.07$ for our initial $|-9/2\rangle$/$|-7/2\rangle$ spin mixture.

At the confinement induced resonance, where $\ln(k_Fa_{2D})=0$, the mean field expansion in powers of $1/\ln(k_Fa_{2D})$ \cite{Bloom1975} breaks down. The energy per particle, after subtracting the two-body bound state energy, approaches a universal value $0.204\,E_F/2$ \cite{Bertaina2010}. Experimentally, we find the frequency shift to cross zero at this point. Note that the gas is very strongly interacting in this regime. According to the optical theorem, the total two-body scattering cross section $\sigma = -\textrm{Im}[f(q)]/q=4/q$ attains the maximal possible value dictated by unitarity, which is essentially the de-Broglie wavelength.

Figure \ref{fig3}b shows the slope of the measured interaction energy vs. magnetic field at the position of the confinement-induced resonance. These data indicate that the interaction energy changes much more smoothly across the confinement-induced resonance as compared to the three-dimensional case (shown at zero lattice depth).

\begin{figure}[htbp]
  \includegraphics[width=0.8\columnwidth,clip=true]{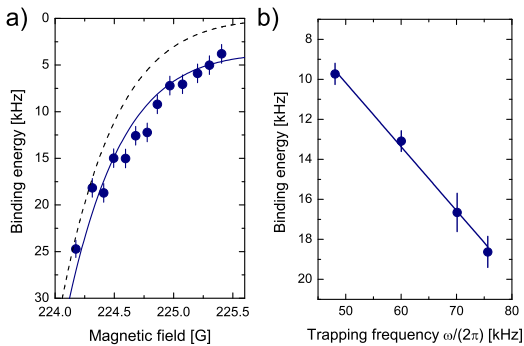}
  \caption{(Color online) a) Binding energy of confinement-induced molecules at a lattice depth of 83\,$E_{rec}$. The dashed line is the prediction of equation (\ref{ebinding}). The solid line is the same curve offset by 4\,kHz to fit the experimental data. b) Binding energy of the confinement-induced molecules at the location of the three-dimensional Feshbach resonance at 224.2\,G. The line is a linear fit to the data. }
  \label{fig4}
\end{figure}

The second peak in the rf spectrum (see Fig. \ref{fig1}) corresponds to confinement-induced bound states. Their existence has been theoretically predicted for an arbitrary value of the three-dimensional scattering length \cite{Petrov2001}. If the frequency of the rf pulse is detuned from the Zeeman transition frequency by the molecular binding energy $E_B$, molecules comprising of atoms in the $|-9/2\rangle$ and $|-5/2\rangle$ states can be associated \cite{Ospelkaus2006}. We demonstrate this on the attractive side of the three-dimensional Feshbach resonance, on which conventional Feshbach molecules do not exist. Upon switching off the optical lattice before the time-of-flight imaging, the confinement-induced molecules are projected into three-dimensions and thus dissociated. As a result, they appear as atoms in the $|-5/2\rangle$ state in absorption imaging. In Fig. \ref{fig4}a we display the measured binding energy as derived from the peak of the molecular spectrum. We observe that the magnitude of the binding energy is larger than the theoretical prediction of the simple two-atom description (eqn. (\ref{ebinding}), dashed line) by 4\,kHz (solid line in Fig. \ref{fig4}a).

The observed energy shift can be attributed to a combination of several effects. One contribution comes from the energy of the relative motion of the atoms before the rf association. In previous experiments \cite{Regal2003b,Moritz2005} this excess kinetic energy was subtracted by determining the minimum energy threshold of the molecular peak. In our spectra the atomic and the molecular peak are partly overlapping for small binding energies of confinement-induced molecules (see Fig. \ref{fig1}) and the extrapolation proves impossible. For a non-interacting, harmonically trapped two-dimensional Fermi gas the kinetic energy contribution amounts to $\langle E\rangle=E_F\left(2/3 + 4 \pi^2 (T/T_F)^2/9+ {\cal{O}}(T^3)\right)$
and the density-weighted average over the different two-dimensional gases results in a shift of $\langle E \rangle=7$\,kHz, which is slightly larger than our observed shift. Another contribution could stem from molecule-molecule and atom-molecule interaction in the final state, the strength of which we estimate below 2\,kHz \cite{Petrov2003}. Moreover, the details of the lineshape, i.e. the relation between the peak in the spectrum and the binding energy, are determined by Franck-Condon factors which so far are calculated only in three dimensions \cite{Chin2005}. Residual deviations between experiment and two-body theory could also hint at many-body pairing effects \cite{Bertaina2010}.

Exactly on the three-dimensional Feshbach resonance, the binding energy of the confinement-induced dimers is predicted to take on the universal value of $E_B=0.244 \hbar \omega$ \cite{Bloch2008}. For these confinement-induced molecules the atomic and the molecular peak are well separated and we have extrapolated the molecular peak in the spectrum linearly to zero kinetic energy. In Fig.~\ref{fig4}b we show this universal scaling of the binding energy as a function of the confinement frequency $\omega$. We observe a linear dependence and find the proportionality constant to be $0.31\pm0.02$, slightly larger than theoretically predicted.

In conclusion, we have investigated a strongly interacting two-dimensional Fermi gas by rf spectroscopy. We have measured the interaction energy shift of the Fermi gas which is dominated by a confinement-induced scattering resonance. Moreover, we have spectroscopically detected two-dimensional confinement-induced molecules. Future studies of superfluid two-dimensional Fermi gases could reveal signatures of the Berezinskii-Kosterlitz-Thouless transition \cite{Petrov2003,Zhang2008} and aid to the understanding of high-temperature superconductivity.

We thank J. Dalibard, Z. Hadzibabic, C. Klempt, C. Kollath, H. Moritz, G. Shlyapnikov for discussions. The work has been supported by {EPSRC} (EP/G029547/1), Daimler-Benz Foundation (B.F.), Studienstiftung, and DAAD (M.F.).

\end{document}